\documentclass[onecolumn,showkeys,preprintnumbers,amsmath,amssymb,prb]{revtex4}
\usepackage{graphicx}
\begin{document}

\title{Quantum transport via evanescent waves in undoped graphene}
\author{M. I. Katsnelson}
\affiliation{Radboud University Nijmegen, Institute for Molecules
and Materials, Heijendaalseweg 135, 6525 AJ, Nijmegen, The
Netherlands }

\begin{abstract}
Charge carriers in graphene are chiral quasiparticles (``massless
Dirac fermions''). Graphene provides therefore an amazing
opportunity to study subtle quantum relativistic effects in
condensed matter experiment. Here I review a theory of one of
these unusual features of graphene, a ``pseudodiffusive''
transport in the limit of zero charge carrier concentration, which
is related to existence of zero-modes of the Dirac operator and to
the Zitterbewegung of unltrarelativistic particles. A conformal
mapping technique is a powerful mathematical tool to study these
phenomena, as demonstrated here, using the Aharonov-Bohm effect in
graphene rings with Corbino geometry as an example.
\end{abstract}
%\pacs{72.80.Vp; 73.23.Ad; 03.65.Pm}

\maketitle

\section{Introduction}
Graphene, a recently discovered two-dimensional allotrope of
carbon, is a subject of a hot scientific interest (for review, see
Refs. \onlinecite{GN07,K07,B08,CNetal08,G09,DSetal10,VKG10}). One
of the most unusual features of this new material is a peculiar
character of electron energy spectrum: the charge carriers in
graphene are chiral quasiparticles described by a two-dimensional
analog of massless Dirac equation with effective ``velocity of
light'' of order of 10$^{6}$ ms$^{-1}$, that is, three hundreds
times smaller than for real massless particles \cite{kostya2,kim}.
This opens a way to study quantum relativistic effects, which were
unreachable in high energy physics, by condensed matter
experiments. As an example, it is worth to mention the theoretical
prediction \cite{KNG06} and experimental observation
\cite{SHG09,YK09} of the Klein tunneling in graphene.

Another amazing property of graphene is the finite minimal
conductivity which is of the order of the conductance quantum
$e^{2}/h$ per valley per spin \cite{kostya2,kim}. Numerous
considerations of the conductivity of a two-dimensional massless
Dirac fermion gas do give this value of the minimal conductivity
with the accuracy of some factor of order of unity
\cite{D1,D2,ludwig,D3,D4,shon,gorbar,D5,K06,Tetal06,Retal07}.

It is really surprising that in the case of massless
two-dimensional Dirac fermions there is a finite conductivity for
an \textit{ideal} crystal, that is, without any scattering
processes \cite{ludwig,K06,Tetal06,Retal07}. This was first
noticed in Ref. \onlinecite{ludwig} using a quite complicated
formalism of conformal field theory (see also a more detailed and
complete discussion in Ref. \onlinecite{Retal07}). After a
discovery of the minimal conductivity in graphene
\cite{kostya2,kim} I was pushed by my colleagues experimentalists
to give a more transparent physical explanation of this fact which
has been done in Ref. \onlinecite{K06}, based on the concept of
Zitterbewegung \cite{S30} and on the Landauer formula
\cite{B91,BB}. The latter approach has been immediately developed
further and used to calculate the shot noise \cite{Tetal06} which
turns out to be similar to that in {\it strongly disordered}
metals (a ``pseudodiffusive transport''). There are now more
theoretical \cite{Petal07,KG08,RRW09,SOGM09,K10} and experimental
\cite{Metal07,Detal08} works studying this regime in a context of
graphene. This situation is very special. For conventional
electron gas in semiconductors, in the absence of disorder, the
states with definite energy (eigenstates of the Hamiltonian) can
be simultaneously the states with definite current (eigenstates of
the current operator) and it is the disorder that results in
non-conservation of the current and finite conductivity. Contrary,
for the Dirac fermions the current operator does not commute with
the Hamiltonian (``Zitterbewegung'') which can be considered as a
kind of intrinsic disorder \cite{K06,AK07}. Therefore, more
detailed understanding of the pseudodiffusive transport in
graphene is not only important for physics of graphene devices but
also has a great general interest for quantum statistical physics
and physical kinetics.

\section{Zitterbewegung as an intrinsic disorder}

The Zitterbewegung is a quantum relativistic phenomenon which was
first discussed by Schr\"{o}dinger as early as in 1930 \cite{S30}.
Only very recently, it was observed experimentally for trapped
ions \cite{Zitexper}. This phenomenon seems to be important to
understand qualitatively peculiarities of electron transport in
graphene at small doping \cite{K06,AK07}. Other aspects of the
Zitterbewegung in graphene physics, in particular, possibilities
of its direct experimental observation, are discussed in Refs.
\cite{Cserti06,Zaw1,Zaw2}. Here we will explain this basic concept
for the case of two-dimensional massless Dirac fermions.

We start with the Hamiltonian
\begin{equation}
H=v\sum\limits_{\mathbf{p}}\Psi _{\mathbf{p}}^{\dagger }\mathbf{\sigma p}%
\Psi _{\mathbf{p}} \label{ham1}
\end{equation}
and the corresponding expression for the current operator
\begin{equation}
\mathbf{j}=ev\sum\limits_{\mathbf{p}}\Psi_{\mathbf{p}}^{\dagger
}\mathbf{\sigma }\Psi_{\mathbf{p}} =
\sum\limits_{\mathbf{p}}\mathbf{j}_{\mathbf{p}} \label{cur1}
\end{equation}
where $v$ is the electron velocity, $\mathbf{\sigma =}\left(
\sigma _{x},\sigma _{y}\right) $ are Pauli matrices, $\mathbf{p}$
is the momentum, and $\Psi _{\mathbf{p}}^{\dagger }=\left( \psi
_{\mathbf{p}1}^{\dagger },\psi _{\mathbf{p}2}^{\dagger }\right) $
\ are pseudospinor electron operators. The expression (\ref{cur1})
follows from Eq.(\ref{ham1}) and the gauge invariance which
requires \cite{abrikos}
\begin{equation}
\mathbf{j}_{\mathbf{p}} =\frac{\delta H}{\delta\mathbf{p}}.
\end{equation}
Here we omit spin and valley indices (so, keeping in mind
applications to graphene, the results for the conductivity should
be multiplied by 4 due to two spin projections and two conical
points per Brillouin zone). Straightforward calculations give for
the time evolution of the electron operators
$\Psi(t)=\exp{(iHt)}\Psi\exp{(-iHt)}$ ($\hbar=1$):
\begin{equation}
\Psi _{\mathbf{p}}\left( t\right) =\frac{1}{2}\left[ e^{-i\epsilon _{\mathbf{%
p}}t}\left( 1+\frac{\mathbf{p\sigma }}{p}\right) +e^{i\epsilon _{\mathbf{p}%
}t}\left( 1-\frac{\mathbf{p\sigma }}{p}\right) \right] \Psi _{\mathbf{p}}
\end{equation}
and for the current operator
\begin{eqnarray}
\mathbf{j}\left( t\right)  &=&\mathbf{j}_{0}\left( t\right) +\mathbf{j}%
_{1}\left( t\right) +\mathbf{j}_{1}^{\dagger }\left( t\right)   \nonumber \\
\mathbf{j}_{0}\left( t\right)  &=&ev\sum\limits_{\mathbf{p}}\Psi _{\mathbf{p}%
}^{\dagger }\frac{\mathbf{p}\left( \mathbf{p\sigma }\right) }{p^{2}}\Psi _{%
\mathbf{p}}  \nonumber \\
\mathbf{j}_{1}\left( t\right)  &=&\frac{ev}{2}\sum\limits_{\mathbf{p}}\Psi _{%
\mathbf{p}}^{\dagger }\left[ \mathbf{\sigma }-\frac{\mathbf{p}\left( \mathbf{%
p\sigma }\right) }{p^{2}}+\frac{i}{p}\mathbf{\sigma \times p}\right] \Psi _{%
\mathbf{p}}e^{2i\epsilon _{\mathbf{p}}t}  \label{current}
\end{eqnarray}
where $\epsilon_{\mathbf{p}}=vp$ is the particle energy. The last
line in Eq.(\ref{current}) corresponds to the Zitterbewegung.

Its physical interpretation is usually given in terms of
Landau-Peierls generalization of the Heisenberg uncertainty
principle \cite{berest,davydov}. Attempts to measure the
coordinate of a relativistic particle with a very high accuracy
requires the energy which is sufficient to create
particle-antiparticle pairs and, thus, we will inevitably loose
our initial particle, being not able to distinguish it from one of
the created ones (according to quantum statistics, all
microparticles are equivalent). This pair creation corresponds to
the oscillating terms with the frequency $2\epsilon _{\mathbf{p}}$
in Eq.(\ref{current}).

In terms of condensed matter physics, the Zitterbewegung is
nothing but a special kind of inter-band transitions with creation
of virtual electron-hole
pairs. The unitary transformation generated by the operator $U_{\mathbf{p}%
}=1/\sqrt{2}(1+i\mathbf{m}_{\mathbf{p}}\mathbf{\sigma )}$, where $\mathbf{m}_{%
\mathbf{p}}=\left( \cos \phi _{\mathbf{p}},-\sin \phi
_{\mathbf{p}}\right) $ and $\phi _{\mathbf{p}}$ is the polar angle
of the vector $\mathbf{p}$, diagonalizes the Hamiltonian and thus
introduces electron and hole states with the energies $\pm vp$;
after this transformation the oscillating term in
Eq.(\ref{current}) corresponds obviously to the inter-band
transitions, e.g.
\begin{equation}
U_{\mathbf{p}}^{\dagger }j_{\mathbf{p}}^{x}U_{\mathbf{p}}=ev\left(
\begin{array}{cc}
-\cos \phi _{\mathbf{p}} & -i\sin \phi _{\mathbf{p}}e^{-i\phi _{\mathbf{p}%
}+2i\epsilon _{\mathbf{p}}t} \\
i\sin \phi _{\mathbf{p}}e^{i\phi _{\mathbf{p}}-2i\epsilon _{\mathbf{p}}t} &
\cos \phi _{\mathbf{p}}
\end{array}
\right) .  \label{current11}
\end{equation}

To calculate the conductivity $\sigma \left( \omega \right) $ one
can try first to use the Kubo formula \cite{zubarev} which reads
for two-dimensional isotropic case:
\begin{equation}
\sigma \left( \omega \right) =\frac{1}{2A}\int\limits_{0}^{\infty
}dte^{i\omega t}\int\limits_{0}^{\beta }d\lambda \left\langle \mathbf{j}%
\left( t-i\lambda \right) \mathbf{j}\right\rangle  \label{kubo11}
\end{equation}
where $\beta =T^{-1}$ is the inverse temperature, $A$ is the
sample area. In the static limit $\omega =0$ taking into account
Onsager relations and analyticity of the correlators $\left\langle
\mathbf{j}\left( z\right) \mathbf{j}\right\rangle$ for $- \beta <
\mathrm{Im}z \leq 0 $ one has \cite{zubarev}
\begin{equation}
\sigma =\frac{\beta }{4A}\int\limits_{-\infty }^{\infty }dt\left\langle
\mathbf{j}\left( t\right) \mathbf{j}\right\rangle .  \label{kubo}
\end{equation}
Usually, for ideal crystals, the current operator commutes with
the Hamiltonian and thus $\mathbf{j}\left( t\right)$ does not
depend on time. In that case, due to Eq.(\ref{kubo11}) the
frequency-dependent conductivity contains only the Drude peak
\begin{equation}
\sigma _{D}\left( \omega \right) =\frac{\pi }{2A}\lim_{T\rightarrow 0}\frac{%
\left\langle \mathbf{j}^{2}\right\rangle }{T}\delta \left( \omega \right)
\label{Drude}
\end{equation}
Either the spectral weight of the Drude peak is finite and, thus,
the static conductivity is infinite, or it is equal to zero. It is
easy to check that for the system under consideration the spectral
weight of the Drude peak is proportional to the modulus of the
chemical potential $\left| \mu \right|$ and thus vanishes at zero
doping ($\mu =0$). It is the Zitterbewegung, i.e. the oscillating
term $\mathbf{j}_{1}\left( t\right) $ which is responsible for
nontrivial behavior of the conductivity for zero temperature and
zero chemical potential. A straightforward calculation gives a
formal result
\begin{equation}
\sigma =\frac{\pi e^{2}}{2h}\int\limits_{0}^{\infty }d\epsilon \epsilon
\delta ^{2}\left( \epsilon \right)  \label{sigma}
\end{equation}
where one delta-function originates from the integration over $t$
in Eq.(\ref {kubo}) and the second one - from the derivative of
the Fermi distribution function appearing at the calculation of
the average over product of Fermi-operators. Of course, the square
of the delta function is not a well-defined object and thus
Eq.(\ref{sigma}) is meaningless before specification of the way
how one should regularize the delta-functions. After
regularization the integral in Eq.(\ref{sigma}) is finite, but its
value depends on the regularization procedure  (for a detailed
discussion of this uncertainty, see Ref. \onlinecite{Retal07}).

Although this derivation cannot give us a correct numerical
factor, it opens a new way to a qualitative understanding of more
complicated situations. For example, the minimal conductivity  of
order of $e^{2}/h$ per channel has been observed experimentally
also for the bilayer graphene \cite{bilayer06} with the energy
spectrum drastically different from that for the single-layer
case. The bilayer graphene is a zero-gap semiconductor with
\textit{parabolic} touching of the electron and hole bands
described by the single-particle Hamiltonian
\cite{bilayer06,falko06}
\begin{equation}
H_{\mathbf{p}}=\left(
\begin{array}{cc}
0 & \left( p_{x}-ip_{y}\right) ^{2}/2m \\
\left( p_{x}+ip_{y}\right) ^{2}/2m & 0
\end{array}
\right)   \label{bilayer}
\end{equation}
(here we ignore some complications due to large-scale hopping
processes which are important for a very narrow range of the Fermi
energies \cite{falko06}). The Hamiltonian (\ref{bilayer}) can be
diagonalized by the unitary
transformation $U_{\mathbf{p}}$ with the replacement $\phi _{\mathbf{p}%
}\rightarrow 2\phi _{\mathbf{p}}.$ Thus, the current operator after the
transformation takes the form (\ref{current11}) with the replacement $%
v\rightarrow p/m,e^{-i\phi_{\mathbf{p}}}\rightarrow e^{-2i\phi _{\mathbf{p}%
}}$. In contrast with the single-layer case, the density of
electron states for the Hamiltonian (\ref{bilayer}) is finite at
zero energy but the square of the current is, vice versa, linear
in energy. As a result, we have the same estimate (\ref{sigma}).

The Zitterbewegung processes play an important role when deriving
quantum kinetic equation for electrons in graphene taking into
account a static disorder \cite{AK07}. For undoped case $\mu =0$
the electron and hole states are degenerate in energy, and
interband (Zitterbewegung) scattering processes are similar, in a
sense, to spin-flip scattering processes in the Kondo problem
\cite{hewson}. As was shown in Ref. \onlinecite{AK07} for a
particular case of short-range scatterers, the Zitterbewegung
processes lead to corrections to the static conductivity which are
logarithmically divergent, and an exponentially small energy scale
similar to the Kondo temperature arises
\begin{equation}
\epsilon_{\text{K}} = \epsilon_{c}e^{-\pi \sigma_{\text{B}}}
\label{qKondo}
\end{equation}
where $\epsilon_c$ the a cutoff energy of order of a bandwidth and
$\sigma_B$ is the conventional Boltzmann conductivity in the units
of $e^2/h$. Earlier this energy scale appeared in nonlinear sigma
model approach to the conductivity of two-dimensional Dirac
fermions \cite{D1,D2}. This makes the classical Boltzmann equation
inapplicable for a very small doping where $|\mu| <
\epsilon_{\text{K}}$.

\section{Landauer formula approach}
A deeper understanding of the origin of finite conductivity
without charge carriers can be reached using Landauer formula
approach \cite{B91}. Following Ref. \onlinecite{K06} we consider
the simplest possible geometry, choosing the sample as a ring of
length $L_{y}$ in $y$ direction; we will use Landauer formula to
calculate the conductance in $x$ direction (see Fig. 1). As we
will see, the conductivity turns out to be dependent on the shape
of the sample. To have a final transparency we should keep $L_{x}$
finite. On the other hand, periodic boundary conditions in $y$
direction are nonphysical and we have to choose $L_{y}$ as large
as possible to weaken their effects. Thus, for two-dimensional
situation one should choose $L_{x}\ll L_{y}.$ In Ref.
\onlinecite{Tetal06} the boundary conditions has been chosen in
two different ways, with metallic armchair edges, and with
infinite gap opening at the edges. The results for the
conductivity in the limit $L_{x}\ll L_{y}$ turned out to be
independent on the boundary conditions.

\begin{figure}[tbp]
\includegraphics[width=9cm]{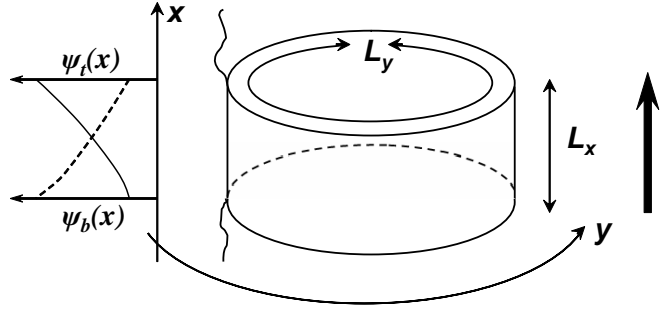}
\caption{Geometry of the sample. Thick arrow shows the direction of current.
$\protect\psi_t$ (solid line) and $\protect\psi_b$ (dashed line) are wave
functions of the edge states localized near the top and the bottom of the
sample, correspondingly. }
\label{fig:1}
\end{figure}

In the coordinate representation the Dirac equation at zero energy takes the
form
\begin{eqnarray}
(K_{x}+iK_{y})\psi _{1} &=&0 \\
(K_{x}-iK_{y})\psi _{2} &=&0  \nonumber
\end{eqnarray}
where $K_{i}=-i\frac{\partial }{\partial x_{i}}.$ General solutions of these
equations are just arbitrary analytical (or complex conjugated analytical)
functions:
\begin{eqnarray}
\psi _{1} &=&\psi _{1}\left( x+iy\right) , \\
\psi _{2}&=&\psi _{2}\left( x-iy\right) .   \nonumber
\label{psi}
\end{eqnarray}
Due to periodicity in $y$ direction both wave functions should be
proportional to $\exp \left( ik_{y}y\right) $ where $k_{y}=2\pi
n/L_{y},n=0,\pm 1,\pm 2,....$ This means that the dependence on
$x$ is also fixed: the wave functions are proportional to $\exp
\left( \pm 2\pi nx/L_{y}\right) .$ They correspond to the states
localized near the bottom and top of the sample (see Fig. 1).

To use the Landauer formula, we should introduce boundary
conditions at the sample edges ($x=0$ and $x=L_{x}$). To be
specific, let us assume that the leads are made of doped graphene
with the potential $V_{0}<0$ and the Fermi energy
$E_{F}=vk_{F}=-V_{0}.$ The wave functions in the leads are
supposed to have the same $y$-dependence, that is, $\psi
_{1,2}\left( x,y\right) =\psi _{t,b}\left( x\right) \exp \left(
ik_{y}y\right) .$ Thus, one can try the solution of the Dirac
equation in the following form:
\begin{eqnarray}
\psi _{t}\left( x\right)  &=&\left\{
\begin{array}{cc}
e^{ik_{x}x}+re^{-ik_{x}x}, & x<0 \\
ae^{k_{y}x}, & 0<x<L_{x} \\
te^{ik_{x}x}, & x>L_{x}
\end{array}
\right.   \nonumber \\
\psi _{b}\left( x\right)  &=&\left\{
\begin{array}{cc}
e^{ik_{x}x+i\phi }-re^{-ik_{x}x-i\phi }, & x<0 \\
be^{-k_{y}x}, & 0<x<L_{x} \\
te^{ik_{x}x+i\phi }, & x>L_{x}
\end{array}
\right.   \label{solution}
\end{eqnarray}
where $\sin \phi =k_{y}/k_{F},k_{x}=\sqrt{k_{F}^{2}-k_{y}^{2}}.$
From the conditions of continuity of the wave functions, one can
find the transmission coefficient
\begin{equation}
T_{n}=\left| t\left( k_{y}\right) \right| ^{2}=\frac{\cos ^{2}\phi
}{\cosh ^{2}(k_{y}L_{x})-\sin ^{2}\phi }.  \label{T1}
\end{equation}
Further, one should assume that $k_{F}L_{x}\gg 1$ and put $\phi
\simeq 0$ in Eq.(\ref{T1}), so,
\begin{equation}
T_n = \frac{1}{\cosh^{2}(k_{y}L_{x})} \label{T}
\end{equation}

The conductance $G$ (per spin per valley) \cite{B91} and Fano
factor of the shot noise \cite{BB} $F$ are expressed via the
transmission coefficients (\ref{T}) as
\begin{eqnarray}
G &=&\frac{e^2}h\sum_{n=-\infty }^\infty T_n,  \nonumber \\
F &=&1-\frac{\sum_{n=-\infty }^\infty T_n^2}{\sum_{n=-\infty
}^\infty T_n} \label{land}
\end{eqnarray}
Note that in the ballistic regime where transmission probability
for a given channel is either one or zero, $F = 0$ (the current is
noiseless) whereas if all $T_n \ll 1$ (e.g., current through
tunnel junctions) $F \approx 1$.

Thus, the trace of the transmission coefficient which is just the
conductance (in units of $e^{2}/h$) is
\begin{equation}
TrT=\sum\limits_{n=-\infty }^{\infty }\frac{1}{\cosh ^{2}(k_{y}L_{x})}\simeq
\frac{L_{y}}{\pi L_{x}}.  \label{T2}
\end{equation}
Assuming that the conductance is equal to $\sigma
\frac{L_{y}}{L_{x}}$ one finds the contribution to the
conductivity equal to $e^{2}/(\pi h)$ \cite{K06,Tetal06}. Note
also that for the case of nanotubes ($L_{x}\gg L_{y}$) one has a
conductance $e^{2}/h$ per channel, in accordance with known
results \cite{tian,chico}.

For the Fano factor one has $F=1/3$ (Ref. \onlinecite{Tetal06}).
This result is very far from the ballistic regime and coincides
with that for strongly disordered metals \cite{BB92,N92}. This
means that, in a sense, the Zitterbewegung works as an intrinsic
disorder.

A similar approach has been applied also to the case of bilayer
graphene \cite{K06b,SB07,Cserti07}. It results in the conductivity
and Fano factor of the same order as those for the single layer.
However, in contrast with the latter case where the results are
robust, for the bilayer numerical factors are dependent on the
order or limits (chemical potential to zero, ratio of the contact
layer thickness to the sample size, etc.) \cite{K06b,SB07}, and
are sensitive to the trigonal warping \cite{Cserti07}, small
deviations from parabolic electron spectrum in the bilayer
graphene \cite{falko06}.

\section{Conformal mapping}

Thus, electron transport in undoped graphene is due to zero modes
of the Dirac operator which are represented by analytic functions
of $z=x+iy$ determined by boundary conditions. For the geometry
shown in Fig. 1 these functions are just exponents:
\begin{equation}
\psi_{1n}(z) = \exp{\left( 2\pi n z/L_y \right)}
\end{equation}
so, a generic wave function inside the graphene flake can be
written as
\begin{equation}
\Psi  ( x , y ) \equiv \sum_{n = - \infty}^{\infty} a_n \left(
\begin{array}{c}
e^{2 \pi n z / L_y} \\
0 \end{array} \right) + b_n \left( \begin{array}{c} 0 \\ e^{2 \pi
n z^{\star} / L_y} \end{array} \right) \label{wf}
\end{equation}
where the coefficients $a_n,b_n$ are determined by the boundary
conditions. Let the Fermi wavelength in the leads much smaller
than geometric sizes of the flake. Then, for the most of modes one
can write the boundary conditions assuming normal incidence,
$\phi=0$:
\begin{eqnarray}
\psi_{in} &\equiv &\left( \begin{array}{c} 1 + r \\ 1 - r
\end{array} \right) \nonumber \\
\psi_{out} &\equiv &\left( \begin{array}{c} t \\ t
\end{array} \right)
\label{rt}
\end{eqnarray}
where subscripts ``in'' and ``out'' label the values of the wave
functions at the boundaries between the leads and the sample. In
this approximation it is very easy to solve the problem of
electron transport through graphene quantum dot of arbitrary shape
using a conformal mapping of this shape to the stripe
\cite{KG08,RRW09}. For example, the mapping
\begin{equation}
w ( z ) = R_1  e^{2 \pi i z / L_y}
\end{equation}
with $e^{2 \pi L_y / L_x} = R_2 / R_1$ transforms the rectangular
stripe $L_x \times L_y$ into the circular ring with inner and
outer radii $R_1$ and $R_2$, respectively. Instead of
Eq.(\ref{wf}) one can try in this case the solutions as
\begin{equation}
\Psi  ( x , y ) \equiv \sum_{n = - \infty}^{\infty} a_n \left(
\begin{array}{c}
z^n \\
0 \end{array} \right) + b_n \left( \begin{array}{c} 0 \\
(z^{\star})^n
 \end{array} \right) \label{wf1}
\end{equation}
The conformal mapping allows us to find immediately the solution
for Corbino geometry where in and out leads are attached to inner
and outer edges of the ring, respectively. Moreover, the solution
of the problem for any shape of the flake topologically equivalent
of the ring can be written automatically in terms of the
corresponding conformal mapping \cite{RRW09}. Earlier \cite{KG08}
this method was applied to the case of graphene quantum dots with
thin leads attached. Following Ref. \onlinecite{K10} we will
present in the next section the results for the Corbino geometry
in a more general case, in the presence of magnetic fields.

\section{Aharonov-Bohm effect in undoped graphene}
The Aharonov-Bohm (AB) effect \cite{AB59,OP85} is the shift of
interference patterns from different electron trajectories by the
magnetic flux through the area between the trajectories. This
leads to oscillations of observable quantities such as conductance
as a function of the magnetic flux. The AB effect in graphene has
been studied already theoretically \cite{Retal07b,Jetal09,Wetal09}
and experimentally \cite{Retal08,Metal09} for the case of a finite
doping. It is not clear {\it a priori} whether this effect is
observable or not in undoped graphene where the transport is
determined by evanescent waves. The analysis \cite{K10} shows
that, whereas for the case of very thin rings the AB oscillations
are exponentially small, for a reasonable ratio of radii, such as,
$R_2/R_1$ =5, the effect is quite observable.

Combining the conformal mapping technique with a general
consideration of zero-energy states for massless Dirac fermions
\cite{AC79} one can derive simple and general rigorous formulas
for any graphene flake topologically equivalent to the ring,
avoiding both numerical simulations and explicit solutions of
Shr\"{o}dinger equation for some particular cases. Note that for
the case of circular ring and constant magnetic field the problem
can be solved exactly for any doping \cite{R10} but, of course,
the mathematics required is much more cumbersome. In the
corresponding limits, the results are, of course, the same.

For the case of zero energy $E=0$ (undoped graphene) the Dirac
equation
\begin{equation}
\mathbf{\sigma }\left( -i\nabla \mathbf{-A}\right) \psi =E\psi
\end{equation}
for the two-component spinor $\psi ,$ $\mathbf{A}$ is the vector
potential and we use the units $\hbar =e=1,$ splits into two
independent equations for the components $\psi _\sigma $ ($\sigma
=\pm $):
\begin{equation}
\left( -i\frac \partial {\partial x}+\sigma \frac \partial
{\partial y}-A_x-i\sigma A_y\right) \psi _\sigma =0  \label{eq}
\end{equation}
We will use the gauge $\nabla \mathbf{A}=0$ and introduce scalar
magnetic potential $\varphi $ by
\begin{equation}
A_x=\frac{\partial \varphi }{\partial y},A_y=-\frac{\partial \varphi }{%
\partial x},
\end{equation}
thus,
\begin{equation}
\nabla ^2\varphi =-B  \label{laplace}
\end{equation}
where $B$ is the magnetic induction. The vector potential can be
eliminated by a substitution \cite{AC79}
\begin{equation}
\psi _\sigma =e^{\sigma \varphi }\chi _\sigma ,  \label{sub}
\end{equation}
Eq.(\ref{eq}) being satisfied for an arbitrary analytic function
$\chi _{+}\left( z \right) $ and complex conjugated analytic
function $\chi _{\_}\left( \overline{z}\right) .$ The latter can
be found from boundary conditions as described in the previous
section. Thus, the boundary conditions following from
Eq.(\ref{rt}) are
\begin{eqnarray}
1+r &=&\psi _{+}^{(1)}  \nonumber   \\
1-r &=&\psi _{-}^{(1)}  \nonumber \\
t &=&\psi _{+}^{(2)}  \nonumber \\
t &=&\psi _{-}^{(2)}  \label{boun} \\
&&  \nonumber
\end{eqnarray}
where superscripts 1 and 2 label the boundaries attached to the
corresponding leads.

If the boundary of the sample is simply connected  one can always
choose $\varphi =0$ at the boundary and, thus, the magnetic field
disappears from Eq.(\ref{boun}); this fact was used in Ref.
\onlinecite{SOGM09} as a very elegant way to prove that a random
vector potential does not change the value of minimal
conductivity. Further we will consider a sample with a topology of
the ring where the scalar potential $\varphi $ is still constant
at each boundary but these constants, $\varphi _1$ and $\varphi
_2$ are different. Also, by symmetry,
\begin{equation}
\chi _{+}^{(2)}/\chi _{+}^{(1)}=\chi _{-}^{(1)}/\chi _{-}^{(2)}
\label{sym}
\end{equation}
The answer for the transmission coefficient $T=\left| t\right| ^2$
has the form:
\begin{equation}
T_j=\frac 1{\cosh ^2\left[ 2\left( j+a\right) \ln \Lambda \right]
} \label{TT}
\end{equation}
where $j=\pm 1/2,\pm 3/2,...$ labels zero modes of the Dirac equation, $%
\Lambda $ is determined by a conformal transformation of our flake
to the rectangle and
\begin{equation}
a=\frac{\varphi _2-\varphi _1}{2\ln \Lambda }  \label{a}
\end{equation}
which generalizes the corresponding result of Ref.
\onlinecite{RRW09} on the case of finite magnetic fields.

For simplicity, we will consider further the case of the Corbino
disc with inner radius $R_1$ and outer radius $R_2$, when
\cite{RRW09}
\begin{equation}
\Lambda =\sqrt{R_2/R_1}  \label{lambda}
\end{equation}
The conductance $G$ (per spin per valley) and Fano factor of the shot noise $%
F$ are expressed via the transmission coefficients (\ref{TT}) via
Eq.(\ref{land}). To calculate the sums one can use the Poisson
summation formula
\begin{equation}
\sum\limits_{n=-\infty }^\infty f\left( n+x\right)
=\sum\limits_{k=-\infty }^\infty \exp \left[ -2i\pi kx\right]
\widetilde{f}_k
\end{equation}
where $n,k$ are all integers and $\widetilde{f}_k$ is the Fourier
transform of the function $f\left( x\right) .$ Substituting
Eq.(\ref{TT}) into (\ref{land}) one finds a compact and general
answer for the effect of magnetic field on the transport
characteristics:
\begin{equation}
G=\frac{e^2}{h\ln \Lambda }\left[ 1+2\sum_{k=1}^\infty \left(
-1\right) ^k\cos \left( 2\pi ka\right) \alpha_k \right]
\label{answ1}
\end{equation}
\begin{equation}
F=1-\frac 23\left[ \frac{1+2\sum_{k=1}^\infty \left( -1\right)
^k\cos \left( 2\pi ka\right) \alpha_k\left( 1+\frac{\pi ^2k^2}{4\ln ^2\Lambda }\right) }{%
1+2\sum_{k=1}^\infty \left( -1\right) ^k\cos \left( 2\pi ka\right)
\alpha_k}\right] \label{answ2}
\end{equation}
where
\begin{equation}
\alpha_k = \frac{\pi ^2k/2\ln \Lambda }{\sinh \left( \pi ^2k/2\ln
\Lambda \right) } \label{alpha}
\end{equation}

Equation (\ref{laplace}) can be solved explicitly for radially
symmetric distributions of the magnetic field $B\left( r\right) $:
\begin{equation}
\varphi _2-\varphi _1=\frac \Phi {2\pi }\ln \left(
\frac{R_2}{R_1}\right)
+\int\limits_{R_1}^{R_2}\frac{dr}r\int\limits_{R_1}^rdr^{\prime
}r^{\prime }B\left( r^{\prime }\right)
\end{equation}
where $\Phi $ is the magnetic flux through the inner ring. In the
case of AB effect where all magnetic flux is concentrated within
the inner ring one has
\begin{equation}
a=\frac{e\Phi }{2\pi \hbar c}  \label{aa}
\end{equation}
where we have restored world constants.

Due to the large factor $\pi^2$ in the argument of sinh in
Eq.(\ref{alpha}) only the terms with $k=1$ should be kept in
Eqs.(\ref{answ1}) and (\ref{answ2}) for all realistic shapes,
thus,
\begin{equation}
G=G_0\left[ 1- \frac{4\pi ^2}{\ln \left( R_2/R_1\right)} \exp
\left( -\frac{\pi ^2}{\ln \left( R_2/R_1\right)} \right) \cos
\left( \frac{e\Phi }{\hbar c}\right)\right] \label{answ3}
\end{equation}
\begin{equation}
F=\frac{1}{3}+\frac{8\pi^4}{3\ln^3\left( R_2/R_1 \right)}
\exp\left(- \frac{\pi^2}{\ln \left( R_2/R_1 \right)} \right) \cos
\left( \frac{e\Phi}{\hbar c} \right)  \label{answ4}
\end{equation}
where
\begin{equation}
G_0=\frac{2e^2}{h\ln(R_2/R_1)}
\end{equation}
is the conductance of the ring without magnetic field
\cite{RRW09}.

Oscillating contributions to $G$ and $F$ are exponentially small
for very thin rings but, for sure, measurable if the ring is thick
enough. For $R_2/R_1=5$ their amplitudes are 5.3\% and 40\%,
respectively.

Consider now a generic case with the magnetic field $B=0$ within
the flake. Then, the solution of Eq.(\ref{laplace}) is a harmonic
function, that is, real or imaginary part by an analytic function.
It can be obtained from the solution for the disc by the same
conformal transformation which determines the function $\Lambda $.
One can see immediately that Eq.(\ref{aa}) remains the same. As
for the expressions (\ref{answ3}), (\ref{answ4})  they can be
rewritten in terms of experimentally measurable quantity $G_0$,
\begin{equation}
G=G_0\left[ 1-\frac{4\pi ^2}{\beta} \exp \left( -\pi ^2/\beta
\right) \cos \left( \frac{e\Phi }{\hbar c}\right)\right]
\label{fine1}
\end{equation}
\begin{equation}
F=\frac 13+ \frac{8\pi ^4}{3\beta ^3}\exp \left( -\pi ^2/\beta
\right) \cos \left( \frac{e\Phi }{\hbar c}\right) \label{fine2}
\end{equation}
where $\beta =2e^2/hG_0$ and we assume $\beta \ll \pi^2$.

Thus, conformal transformations \cite{KG08,RRW09} is a powerful
tool to consider pseudodiffusive transport in undoped graphene
flakes of arbitrary shape, not only without magnetic field but
also in the presence of magnetic fluxes in the system.
Experimental study of the Aharonov-Bohm oscillations and
comparison with simple expressions (\ref{fine1}), (\ref{fine2})
derived here would be a suitable way to check whether the
ballistic (pseudodiffusive) regime is reached or not in a given
experimental situation.

\section{Conclusions}
Undoped graphene is a gapless semiconductor, with filled valence
band and empty conduction band. It is really counterintuitive that
in such a situation, at zero temperature, it has a finite
conductivity, of order of the conductance quantum $e^2/h$. This is
one of the most striking consequences of its peculiar
``ultrarelativistic'' energy spectrum. More precisely, this is the
consequence of {\it chirality} of electron states. In the case of
bilayer \cite{bilayer06} the electron energy spectrum is parabolic
but the states are chiral, with the Berry phase $2\pi$, and this
leads to the finite minimal conductivity \cite{K06b,SB07,Cserti07}
as discussed above.

Formally, the electron transport in undoped graphene is determined
by zero modes of the Dirac operator which are described by
analytic functions with proper boundary conditions. Therefore, the
whole power of complex calculus can be used here, as well as in
classical old-fashionable branches of mathematical physics such as
two-dimensional hydrodynamics or electrostatics. These states
cannot correspond to the waves propagating through the sample but,
rather, are represented by evanescent waves. The transport via
evanescent waves in undoped graphene is a completely new variety
of electron transport in solids, drastically different from all
types known before (ballistic transport in nanowires and
constrictions, diffusive transport in dirty metals,
variable-range-hopping transport in Anderson insulators, etc.).
Deeper understanding of this new quantum phenomena seems to be a
very important problem.

The work is financially supported by Stichting voor Fundamenteel
Onderzoek der Materie (FOM), the Netherlands.


\begin{thebibliography}{99}

\bibitem{GN07} A. K. Geim and K. S. Novoselov, Nature Mater. {\bf
6},183 (2007).

\bibitem{K07} M. I. Katsnelson, Mater. Today {\bf 10}, 20 (2007).

\bibitem{B08} C. W. J. Beenakker, Rev. Mod. Phys. {\bf 80}, 1337
(2008).

\bibitem{CNetal08} A. H. Castro Neto, F. Guinea, N. M. R. Peres,
K. S. Novoselov, and A. K. Geim, Rev. Mod. Phys. {\bf 80}, 315
(2008).

\bibitem{G09} A. K. Geim, Science {\bf 324}, 1530 (2009).

\bibitem{DSetal10} S. Das Sarma, S. Adam, E. H. Hwang, and E.
Rossi, arXiv:1003.4731.

\bibitem{VKG10} M. A. H. Vozmediano, M. I. Katsnelson, and F.
Guinea, Phys. Rep. {\bf 496}, 109 (2010).

\bibitem{kostya2}  K. S. Novoselov, A. K. Geim, S. V. Morozov, D. Jiang, M.
I. Katsnelson, I. V. Grigorieva, S. V. Dubonos, and A. A. Firsov,
Nature \textbf{438}, 197 (2005).

\bibitem{kim}  Y. Zhang, Y.-W. Tan, H. L. Stormer, and P. Kim, Nature
\textbf{438}, 201 (2005).

\bibitem{KNG06} M. I. Katsnelson, K. S. Novoselov, and A. K. Geim, Nature Phys. {\bf 2}, 620 (2006).

\bibitem{SHG09} N. Stander, B. Huard, and D. Goldhaber-Gordon, Phys. Rev. Lett. {\bf 102}, 026807
(2009).

\bibitem{YK09} A. F. Young and P. Kim, Nature Phys. {\bf 5}, 222 (2009).

\bibitem{D1}  E. Fradkin, Phys. Rev. B \textbf{33}, 3263 (1986).

\bibitem{D2}  P. A. Lee, Phys. Rev. Lett. \textbf{71}, 1887 (1993).

\bibitem{ludwig}  A. W. W. Ludwig, M. P. A. Fisher, R. Shankar, and G.
Grinstein, Phys. Rev. B \textbf{50}, 7526 (1994).

\bibitem{D3}  A. A. Nersesyan, A. M. Tsvelik, and F. Wenger, Phys. Rev.
Lett. \textbf{72}, 2628 (1994).

\bibitem{D4}  K. Ziegler, Phys. Rev. Lett. \textbf{80}, 3113 (1998).

\bibitem{shon}  N. H. Shon and T. Ando, J. Phys. Soc. Japan \textbf{67},
2421 (1998).

\bibitem{gorbar}  E. V. Gorbar, V. P. Gusynin, V. A. Miransky, and I. A.
Shovkovy, Phys. Rev. B \textbf{66}, 045108 (2002).

\bibitem{D5}  X. Yang and C. Nayak, Phys. Rev. B \textbf{65}, 064523 (2002).

\bibitem{K06} M. I. Katsnelson, Eur. Phys. J. B {\bf 51}, 157
(2006).

\bibitem{Tetal06} J. Tworzydlo, B. Trauzettel, M. Titov, A. Rycerz, and C. W. J.
Beenakker, Phys. Rev. Lett. {\bf 96}, 246802 (2006).

\bibitem{Retal07} S. Ryu, C. Mudry, A. Furusaki, and A. W. W. Ludwig, Phys. Rev. B 75, 205344
(2007).

\bibitem{S30} E. Schr\"{o}dinger, Sitz. Preuss. Akad. Wiss. Phys.-Math. {\bf 24}, 418 (1930).

\bibitem{B91}  C. W. J. Beenakker and H. van Houten, Solid State Phys.
\textbf{44}, 1 (1991).

\bibitem{BB} Ya. M. Blanter and M. B\"{u}ttiker, Phys. Rep. {\bf 336}, 1
(2000).

\bibitem{Petal07} E. Prada, P. San-Jose, B. Wunsch, and F. Guinea,
Phys. Rev. B {\bf 75}, 113407 (2007).

\bibitem{KG08} M. I. Katsnelson and F. Guinea, Phys. Rev. B,
{\bf 78}, 075417 (2008).

\bibitem{RRW09} A. Rycerz, P. Recher, and M. Wimmer, Phys.
Rev. B {\bf 80} 125417 (2009).

\bibitem{SOGM09} A. Schuessler, P. M. Ostrovsky, I. V. Gornyi, and
A. D. Mirlin, Phys. Rev. B {\bf 79}, 075405 (2009).

\bibitem{K10} M. I. Katsnelson, Europhys. Lett. {\bf 89}, 17001
(2010).

\bibitem{Metal07} F. Miao, S. Wijeratne, Y. Zhang, U. C. Coskun, W. Bao, and
C. N. Lau, Science {\bf 317}, 1530 (2007).

\bibitem{Detal08} R. Danneau, F. Wu, M. F.Craciun, S. Russo, M. Y. Tomi,
J. Salmilehto, A. F. Morpurgo, and P. J. Hakonen, Phys. Rev. Lett. {\bf 100}, 196802 (2008).

\bibitem{AK07} M. Auslender and M. I. Katsnelson, Phys. Rev.
B {\bf 76}, 235425 (2007).

\bibitem{Zitexper} R. Gerritsma, G. Kirchmair, F. Z\"{a}hringer, E. Solano, R. Blatt, and C. F.
Roos, Nature {\bf 463}, 68 (2010).

\bibitem{Cserti06} J. Cserti and G. D\'{a}vid, Phys. Rev. B {\bf 74}, 172305
(2006).

\bibitem{Zaw1} T. M. Rusin and W. Zawadzki, Phys. Rev. B {\bf 78}, 125419
(2008).

\bibitem{Zaw2} T. M. Rusin and W. Zawadzki, Phys. Rev. B {\bf 80}, 045416
(2009).

\bibitem{abrikos}  A. A. Abrikosov, Phys. Rev. B {\bf 58}, 2788 (1998).

\bibitem{berest}  V. B. Berestetskii, E. M. Lifshitz, and L. P. Pitaevskii,
{\it Relativistic Quantum Theory}, vol. 1 (Pergamon, Oxford etc.,
1971).

\bibitem{davydov}  A. S. Davydov, {\it Quantum Mechanics} (Pergamon,
Oxford etc., 1976).

\bibitem{zubarev}  D. N. Zubarev, {\it Nonequilibrium Statistical
Thermodynamics} (Consultants Bureau, New York, 1974).

\bibitem{bilayer06}  K. S. Novoselov, E. McCann, S. V. Morozov, V. I. Falko,
M. I. Katsnelson, U. Zeitler, D. Jiang, F. Schedin, and A. K.
Geim, Nature Phys. {\bf 2}, 177 (2006).

\bibitem{falko06}  E. McCann and V. I. Falko, Phys. Rev. Lett.
{\bf 96}, 086805 (2006).

\bibitem{hewson} A. C. Hewson, {\it The Kondo Problem to Heavy
Fermions} (Cambridge University Press, Cambridge, 1993).

\bibitem{tian}  W. Tian and S. Datta, Phys. Rev. B {\bf 49}, 5097 (1994).

\bibitem{chico}  L. Chico, L. X. Benedict, S. G. Louie, and M. L. Cohen,
Phys. Rev. B  {\bf 54}, 2600 (1996).

\bibitem{BB92} C. W. J. Beenakker and M. B\"{u}ttiker, Phys. Rev. B {\bf 46},
1889 (1992).

\bibitem{N92} K. E. Nagaev, Phys. Lett. A {\bf 169}, 103 (1992).

\bibitem{K06b} M. I. Katsnelson, Eur. Phys. J. B {\bf 52}, 151
(2006).

\bibitem{SB07} I. Snyman and C. W. J. Beenakker, Phys. Rev. B {\bf 75}, 045322
(2007).

\bibitem{Cserti07} J. Cserti, A. Csord\'{a}s, and G. D\'{a}vid, Phys. Rev. Lett. {\bf 99}, 066802
(2007).

\bibitem{AB59} Y. Aharonov and D. Bohm, Phys. Rev. {\bf 115}, 485 (1959).

\bibitem{OP85} S. Olariu andI. I. Popescu, Rev. Mod. Phys.
{\bf 57}, 339 (1985).

\bibitem{Retal07b} P. Recher, B. Trauzettel, A. Rycerz, Ya.
M. Blanter, C. W. J. Beenakker, and A. F. Morpurgo, Phys. Rev. B
{\bf 76}, 235404 (2007).

\bibitem{Jetal09} R. Jackiw, A. I. Milstein, S.-Y. Pi, and
I. S. Terekhov, Phys. Rev. B {\bf 80}, 033413 (2009).

\bibitem{Wetal09} J. Wurm, M. Wimmer, H. U. Baranger, and K. Richter, Semicond. Sci.
Tech. {\bf 25}, 034003 (2010).

\bibitem{Retal08} S. Russo, J. B. Oostinga, D. Wehenkel, H. B. Heersche, S. S. Sobhani,
L. M. K. Vandersypen, and A. F. Morpurgo, Phys. Rev. B {\bf 77},
085413 (2008).

\bibitem{Metal09} M. Huefner, F. Molitor,  A. Jacobsen, A. Pioda,
C. Stampfer, K. Ensslin, and T. Ihn, Phys. Stat. Sol. (b) {\bf
246}, 2756 (2009).

\bibitem{AC79} Y. Aharonov and A. Casher, Phys. Rev. A {\bf 19},
2461 (1979).

\bibitem{R10} A. Rycerz, Phys. Rev. B {\bf 81}, 121404 (2010).

%\bibitem{FGM08} M. Fogler, F. Guinea, and  M. I. Katsnelson, Phys.
%Rev. Lett. {\bf 101}, 226804 (2008).

%\bibitem{GKG09} F. Guinea, M. I. Katsnelson, and A. K. Geim,
%Nature Phys. {\bf 6}, 30 (2009).
\end{thebibliography}
\end{document}